\documentclass[3p,preprint]{elsarticle}

\usepackage{amssymb}
\usepackage{amsmath} % to use aligned environment

\usepackage{lineno}

\usepackage{amssymb,amsthm,MnSymbol}
\usepackage{color}

\usepackage{algorithm}
\usepackage{algorithmic}

\usepackage{subfig}

\usepackage{amscd}
\usepackage{mathrsfs}
\usepackage[colorlinks]{hyperref}
\usepackage{enumitem}
\definecolor{darkgreen}{rgb}{0.0,0.75,0.0}              % Light green
\definecolor{lightblue}{rgb}{0.3296, 0.6648, 0.8644}    % Sky blue
\definecolor{shadowcolor}{rgb}{0.0000, 0.0000, 0.6179}  % Midnight blue
\definecolor{bulletcolor}{rgb}{ 0.8441, 0.1582, 0.0000} % Orange-red
\definecolor{deepskyblue}{rgb}{0, 0.75, 1.0}
\definecolor{royalblue}{rgb}{0.254901960784314,   0.411764705882353,   0.882352941176471}
\definecolor{dodgerblue}{rgb}{0.11765, 0.56471, 1.0}
\definecolor{bordercolor}{rgb}{0,0,.2380}               % Midnight blue

\definecolor{lightskyblue}{rgb}{0.53, 0.81, 0.98}
\definecolor{magenta}{rgb}{1.0, 0.0, 1.0}
\definecolor{lavendermagenta}{rgb}{0.93, 0.51, 0.93}
\definecolor{internationalorange}{rgb}{1.0, 0.31, 0.0}

\newcommand{\ansA}[1]{{\color{black}#1}}
\newcommand{\ansB}[1]{{\color{black}#1}}
\newcommand{\ansC}[1]{{\color{black}#1}}
\newcommand{\ansD}[1]{{\color{black}#1}}

\usepackage[normalem]{ulem}

\renewcommand\sout{\bgroup\protect\markoverwith{\textcolor{black}{\protect\rule[0.5ex]{2pt}{0.8pt}}}\protect\ULon}

\biboptions{authoryear}

\usepackage[figuresright]{rotating}

\usepackage{epstopdf} % ???Igor

\begin{document}

\begin{frontmatter}



% \author[addr1]{}
\author{I. Shevchenko\corref{cor1}}
\ead{i.shevchenko@imperial.ac.uk}
\author{P. Berloff}
 
%\fntext[label2]{}
\cortext[cor1]{Corresponding author at:}
%\address{Address\fnref{label3}}
%% \fntext[labecl3]{}

\address{Department of Mathematics, Imperial College London, Huxley Building, 180 Queen's Gate, London, SW7 2AZ, UK.}
% \address[addr1]{Department of Mathematics, Imperial College London, Huxley Building, 180 Queen's Gate, London, SW7 2AZ, UK.}

% \dochead{Short communication}
%% e.g. \dochead{17th International Conference on Dynamical Processes in Excited States of Solids}

\title{On a minimum set of equations for parameterisations in comprehensive ocean circulation models}


%\author{}

%\address{}

\begin{abstract}
% \linenumbers
% \onehalfspacing % ???
The complexity of comprehensive ocean models poses an important question for parameterisations: is there a minimum set of equations that should be parameterised, on the one hand, to reduce the development to a minimum, and, on the other hand, to ensure an accurate representation of large-scale flow patterns?
This work seeks to answer this to assist modern parameterisations be more selective in their targets.
For this, we considered a North Atlantic configuration of the MIT general circulation model and studied contributions of different model equations to the accuracy of representation of the Gulf Stream at low resolution.
Our results suggest that it is enough to parameterise only the tracer equations for temperature and salinity and leave the other equations in the hydrodynamic part, as well as the atmospheric model unmodified.
\end{abstract}

\begin{keyword}
Ocean general circulation models \sep Eddy parameterisations \sep Geophysical fluid dynamics
\end{keyword}

\end{frontmatter}

% \linenumbers

% \clearpage
% \tableofcontents

\section{Introduction}
The development of parameterisations for comprehensive ocean models encounters significant challenges not only because of the chaotic nature of the flow but also because of the complexity of the models themselves.
Typically, modern ocean models are coupled with atmospheric models and consist of many equations.
This raises an important question: Which equations should be parameterised first of all?
Many modern parameterisations for both comprehensive and idealized ocean models 
(e.g.,~\citet{
DuanNadiga2007,Frederiksen_et_al2012,JansenHeld2014,
PortaMana_Zanna2014,CooperZanna2015,
Grooms_et_al2015,Berloff_2015,Berloff_2016,Berloff_2018,
Danilov_etal_2019,Ryzhov_etal_2019,JAKK_2019,Bachman_2019,
Juricke_etal_2020a,Juricke_etal_2020b,
CCHWS2019_1,Ryzhov_etal_2020,
CCHWS2019_3,CCHWS2020_4,CCHPS2020_J2})
target the momentum equation and try to fix the kinetic energy backscatter mechanism to ensure energy-consistent parameterisations, while avoiding parameterisations in the tracer equations (temperature and salinity).
The Gent-McWilliams parameterisation~\citep{GentMcwilliams1990}, on the contrary, aims at reproducing eddy effects of baroclinic instability by flattening the isopycnals and deducing the corresponding bolus fluxes to be taken into account for transport of mass and tracers.
\ansD{Note that typically the range of applicability of the eddy backscatter parameteriation and Gent-McWilliams parameterisation is different.
The former addresses eddy-permitting models while the latter addresses non-eddy resolving models.}

In this paper we show that parameterising only the momentum part leads to an insurmountable degradation of the large-scale fields in the tracer equations and therefore requires an additional parameterisation for the latter.
\ansB{On the other hand}, a parameterisation only in the tracer equations may be enough to preserve large-scale flow structures.

%
%%%%%%%%%%%%%%%%%%%% MODEL CONFIGURATION
%
\section{Model configuration and augmentation}
\begin{figure}[H]
\centering
\hspace*{-3.5cm}
\begin{tabular}{c}
\hspace*{4.5cm}\begin{minipage}{0.1\textwidth} {\bf (a)} \end{minipage} 
\hspace*{3.7cm}\begin{minipage}{0.1\textwidth} {\bf (b)} \end{minipage} 
\hspace*{3.7cm}\begin{minipage}{0.1\textwidth} {\bf (c)} \end{minipage}\hspace*{-0.25cm}\\
\hspace*{0cm}\begin{minipage}{0.8\textwidth}\includegraphics[scale=1.25]{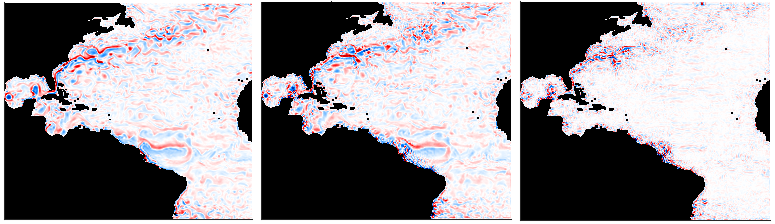}\end{minipage}\\
\hspace*{0.1cm}\begin{minipage}{0.8\textwidth}\includegraphics[scale=1.25]{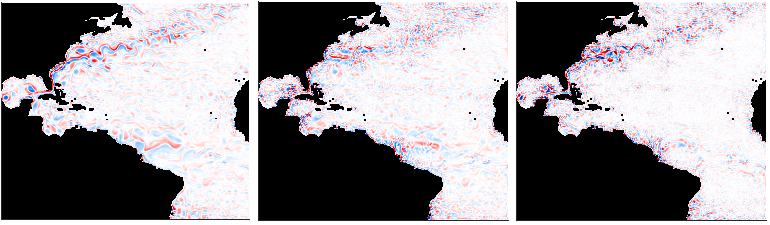}\end{minipage}\\
\hspace*{3.5cm}\begin{minipage}{0.35\textwidth}\includegraphics[scale=1.25]{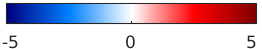}\end{minipage}\\
\end{tabular}
\caption{
Instantaneous surface relative vorticity $\zeta=v_x-u_y$ \rm[1/s] for {\bf (a)} the truth, {\bf (b)} coarse solution started from the true initial condition, {\bf (c)} the difference between (a) and (b).
Snapshots are taken after (top panels) 7 and (bottom panels) 30 days of simulations.
}
\label{fig:rv_12_3}
\end{figure}

We use the Massachusetts Institute of Technology general circulation model (MITgcm)~\citet{Marshall_etal_1997} with the North Atlantic configuration similar to the one in~\citep{Jamet_etal_2019}. 
\ansD{At the surface, the oceanic model is coupled with an atmospheric model~\citep{Marshall_etal_1997,Deremble_et_al2013}.}
% The only difference is the computational domain, which in our case is $[98^\circ \rm W,12^\circ \rm W]\times[20^\circ \rm S,55^\circ \rm N]$. 
We integrate the \ansD{coupled} model at two different horizontal resolutions, namely $1/12^{\circ}$ and $1/3^{\circ}$, and refer to the higher-resolution solution projected onto the coarse grid (Fig.~\ref{fig:rv_12_3}a) as the ``truth'' (or ``true solution''), and to the solution computed on the coarser grid as the coarse solution (Fig.~\ref{fig:rv_12_3}b).
The model was initially spun up from the state of rest over the time interval of 5 years.
Formally, the initial condition for all our subsequent solutions is the truth at midnight on 01 January 1963.
Note that for the purposes of this study it is not essential that the initial state of the ocean circulation is in the statistically equilibrated regime.

As seen in Fig.~\ref{fig:rv_12_3}, the coarse-grid model fails to preserve the large-scale flow pattern of the Gulf Stream, and especially of its eastward jet extension. 
Namely, the Gulf Stream is gradually degraded, and nearly vanishes after only 30 days; the same is observed for temperature and salinity (not shown).
This situation is typical for low-resolution solutions and usually attributed to the inability of the coarse-gridded numerical equations to capture effects of the unresolved small-scale dynamics on the resolved large scales \ansC{as well as excessive numerical damping and mixing}.
The small-scale effects potentially can \ansC{potentially} be parameterised, but which of the involved equations are to be prioritized in terms of parameterisations, given that we are interested in both minimizing computational costs and achieving reasonably accurate representation of large-scale flow patterns?
In order to answer this question, we augmented different equations of the coarse-grid model and studied how this affected the large-scale flow patterns 
(with the focus on the Gulf Stream extension; not to be mentioned again).

The idea behind the implemented augmentation is to correct the coarse solution towards the true one using the available ``true-solution'' information (e.g.~\citet{Berloff_2005b,Berloff_2005a}).
The corrected (i.e., augmented) solution $\phi_A$ is computed as follows:
\begin{equation}
\phi_A:=\overline{\phi}+\eta(\phi-\overline{\phi}) \, ,
\label{eq:phi_a}
\end{equation}
where $\overline{\phi}$ is the coarse solution, $\phi$ is the true solution, and $\eta$ is the augmentation amplitude, and the augmentation is performed at the end of every augmentation step (6 hours). 
\ansA{Note that we augment the full solution to the corresponding equation (not particular terms in the equation).} 
\ansC{In the oceanic model we augment the solutions to the momentum equation and tracer equations. 
In the atmospheric model, atmosphere boundary layer temperature and atmosphere specific humidity are augmented.}
Occasionally, we use the term ``augment an equation'' to mean ``augment the solution of that equation''.  
In fact, augmentation can be considered as the perfect parameterisation that uses all available information and produces a coarse solution matching up the truth.

%
%%%%%%%%%%%%%%%%%%%% NUMERICAL RESULTS
%
\section{Numerical results}
Our goal is to maintain the Gulf Stream in the coarse-grid model by augmenting a minimal set of equations.
For this, we carried out a series of simulations and studied how augmentation of different equations in the model influences the solution.
First, we only augmented the momentum equation and the atmospheric model, and looked at how this affected other variables, such as temperature, salinity, and sea surface height (Fig.~\ref{fig:T_S_Eta}); 
\ansC{note that both the oceanic and atmospheric models are implemented with the same resolution.}

\begin{figure}[H]
\centering
\hspace*{-3.0cm}
\subfloat{
\begin{tabular}{c}
\hspace*{3.5cm}\begin{minipage}{0.1\textwidth} {\bf (a)} \end{minipage} 
\hspace*{3.7cm}\begin{minipage}{0.1\textwidth} {\bf (b)} \end{minipage} 
\hspace*{3.7cm}\begin{minipage}{0.1\textwidth} {\bf (c)} \end{minipage}\hspace*{-0.25cm}\\
\hspace*{2.5cm} {\bf Sea surface temperature} \\
\hspace*{0cm}\begin{minipage}{0.8\textwidth}\includegraphics[scale=1.2]{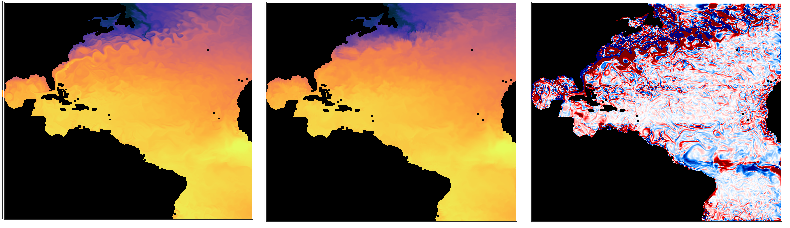}\end{minipage}\\
\hspace*{0cm}\begin{minipage}{0.8\textwidth}\includegraphics[scale=1.2]{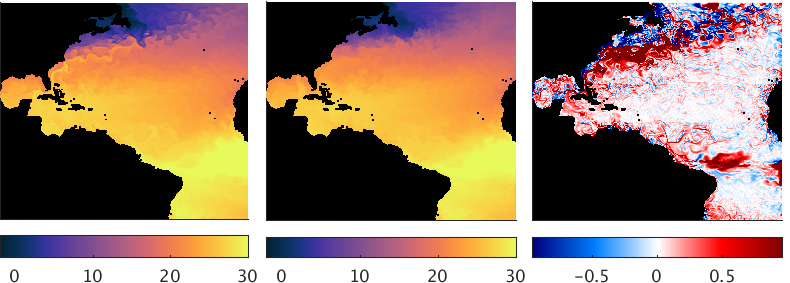}\end{minipage}\\
\\[-0.25cm]
\hspace*{2.75cm} {\bf Sea surface salinity} \\
\hspace*{-0.5cm}\begin{minipage}{0.8\textwidth}\includegraphics[scale=1.2]{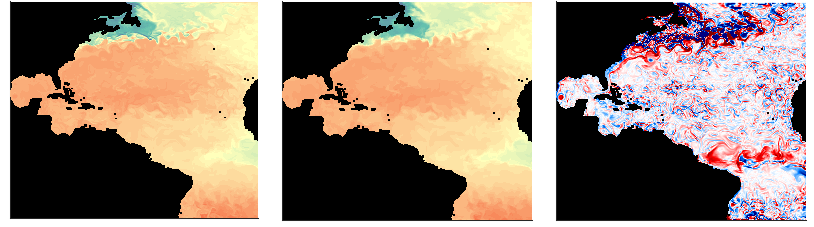}\end{minipage}\\
\hspace*{-0.5cm}\begin{minipage}{0.8\textwidth}\includegraphics[scale=1.2]{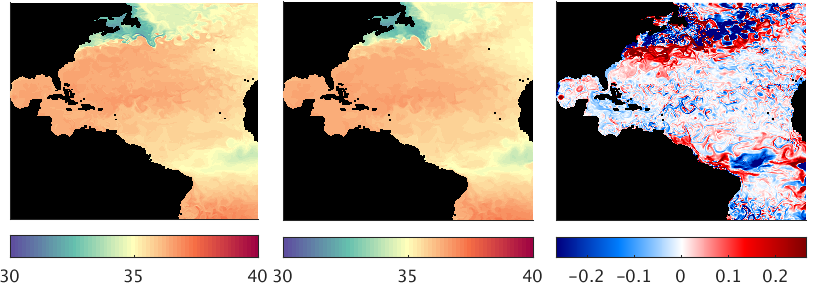}\end{minipage}\\
\end{tabular}
}
\caption{Figure continued on next page.}
\label{fig:T_S_Eta}
\end{figure}
\setcounter{figure}{1}    
\begin{figure}[H]
\subfloat{
\hspace*{-0.35cm}
\begin{tabular}{c}
\hspace*{2.75cm} {\bf Sea surface height} \\
\hspace*{0cm}\begin{minipage}{0.8\textwidth}\includegraphics[scale=1.25]{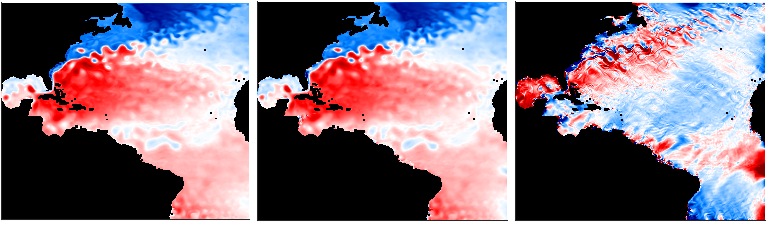}\end{minipage}\\
\hspace*{0cm}\begin{minipage}{0.8\textwidth}\includegraphics[scale=1.25]{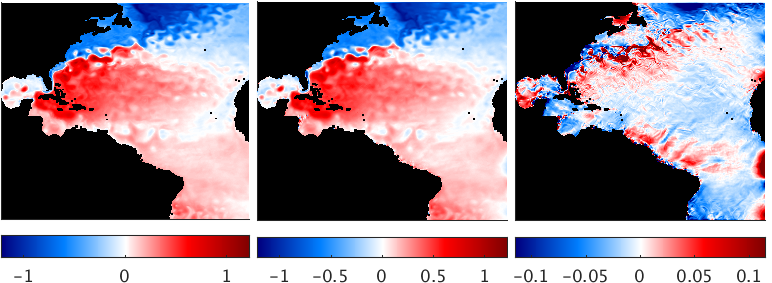}\end{minipage}\\
\end{tabular}}
\caption{
Shown is an instantaneous sea surface temperature [$^\circ{\rm C}$], sea surface salinity [{\rm g/kg}], and sea surface height [{\rm m}] for {\bf (a)} the truth, {\bf (b)} coarse solution \ansC{(with the augmented momentum equation and atmospheric model, the augmentation amplitude is $\eta=1$)} started from the true initial condition, {\bf (c)} the difference between (a) and (b); snapshots are taken after (top panels) 1 and (bottom panels) 3 months of simulations.
}
\label{fig:T_S_Eta}
\end{figure}

\begin{figure}[H]
\centering
\hspace*{-3.25cm}
\begin{tabular}{c}
\hspace*{3.5cm}\begin{minipage}{0.1\textwidth} {\bf (a)} \end{minipage} 
\hspace*{3.7cm}\begin{minipage}{0.1\textwidth} {\bf (b)} \end{minipage} 
\hspace*{3.7cm}\begin{minipage}{0.1\textwidth} {\bf (c)} \end{minipage}\hspace*{-0.25cm}\\
\hspace*{2.5cm} {\bf Augmentation of temperature and salinity in the whole domain} \\
\hspace*{0cm}\begin{minipage}{0.8\textwidth}\includegraphics[scale=1.25]{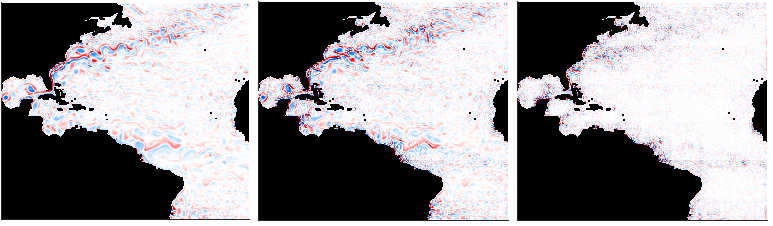}\end{minipage}\\
\hspace*{0cm}\begin{minipage}{0.8\textwidth}\includegraphics[scale=1.25]{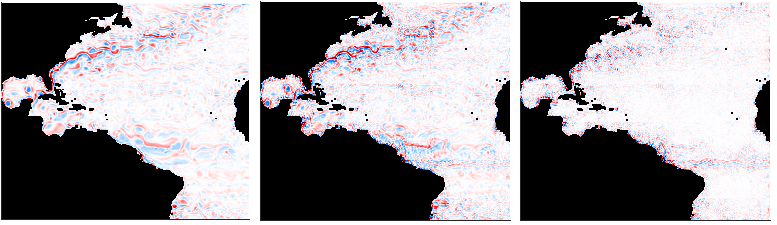}\end{minipage}\\
\\
\hspace*{3cm} {\bf Augmentation of temperature and salinity only in the Gulf Stream region (red rectangular)} \\
\hspace*{0cm}\begin{minipage}{0.8\textwidth}\includegraphics[scale=1.25]{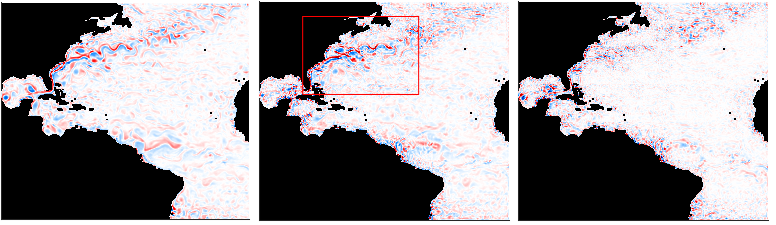}\end{minipage}\\
\hspace*{0cm}\begin{minipage}{0.8\textwidth}\includegraphics[scale=1.25]{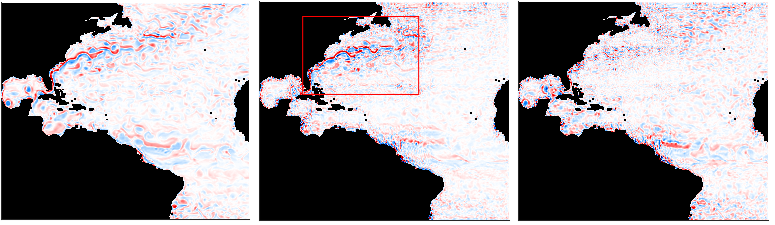}\end{minipage}\\
\hspace*{3.5cm}\begin{minipage}{0.35\textwidth}\includegraphics[scale=1.25]{colorbar.png}\end{minipage}\\
\end{tabular}
\caption{
Instantaneous surface relative vorticity $\zeta=v_x-u_y$ \rm[1/s] for {\bf (a)} the truth, {\bf (b)} coarse solution with augmentation of temperature and salinity \ansC{(the augmentation amplitude is $\eta=1$)}, {\bf (c)} the difference between (a) and (b).
Snapshots are taken after (top panels) 1 and (bottom panels) 12 months of simulations.
From comparison, we found that the augmentation can be local (only in the Gulf Stream region) but still successful.
}
\label{fig:augment_T_S__no_augment_u_v_eta_Tair_Qair}
\end{figure}

As seen in Fig.~\ref{fig:T_S_Eta}, the augmentation of only the momentum equation and the atmospheric model results in significant degradation of temperature and salinity. 
\ansD{Although the Gulf Stream 
separation point is still in place, the entire flow is inhibited. Namely, the Gulf Stream in the augmented model 
is shorter and weaker (it is seen in the temperature and salinity fields) compared to the reference solution.}
It suggests that the corresponding individual parameterisations would not be sufficient
At the same time the sea surface height equation can be left unparameterised, because its augmentation produces insignificant changes.
However, we found that the coarse model can maintain the Gulf Stream with augmentation of only the tracer equations (Fig.~\ref{fig:augment_T_S__no_augment_u_v_eta_Tair_Qair}), and there is no need to augment the other equations.
Moreover, this augmentation can be implemented only locally in the Gulf Stream region 
(Fig.~\ref{fig:augment_T_S__no_augment_u_v_eta_Tair_Qair}), and the result is qualitatively the same.

We also studied effects of augmenting each tracer equation separately and found that it leads to very substantial and fast (on the scale of a month) degradation of the solution.
In this case the solution becomes quickly contaminated with high-frequency waves which ruin its large-scale structure.
Therefore, we conclude that both equations are to be augmented simultaneously, in order to have the density field augmented correctly.
\ansC{This can be used as a general guidance on how to shape up a good parameterisation.}

% Basically, in the discussed scenario we pump in EPE, thus, energizing the EPE backscatter.
% However, part of this energy has to be converted into KE (on different scales), and these conversion mechanisms
% are not adequately represented. In this sense we agree with the reviewer's statement that "the model is missing
% more realistic dissipation mechanisms". To dig into this further, we'll have to separate eddies from the large
% scales (by some filtering) and carry out complete analysis of the energetics --- this is doable but very difficult
% task (we already have relevant experiences with energetics analyses).

\begin{figure}[H]
\centering
\hspace*{0cm}\begin{minipage}{1.0\textwidth}\includegraphics[scale=0.26]{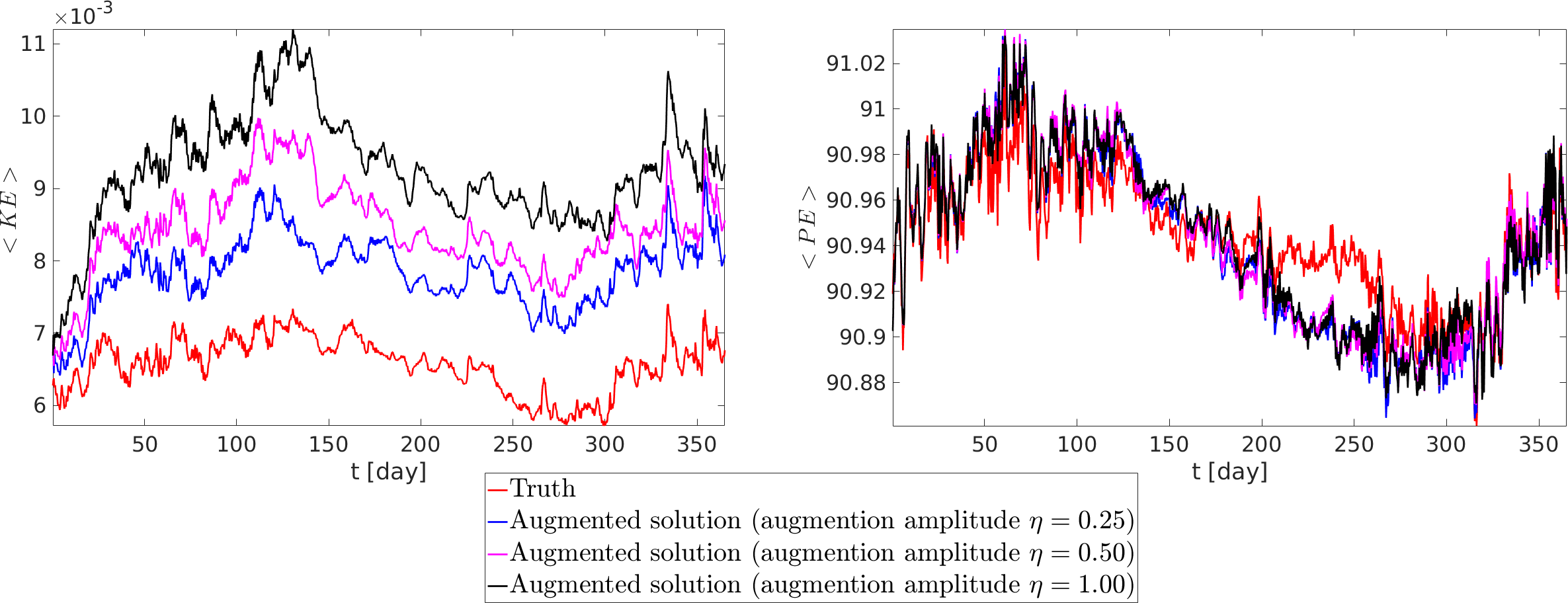}\end{minipage}
\caption{
Augmentation of the temperature and salinity equations in the Gulf Stream region.
Dependence of 800-meter depth-averaged kinetic (left) and potential (right) energies on the augmentation amplitude $\eta$; units are $\rm[m^2/s^2]$. 
}
\label{fig:KP_energy_GSonly_amplitude}
\end{figure}

Let us now focus on the augmentation of both tracer equations simultaneously.
The analysis of kinetic and potential energies (KE and PE, respectively) shows that the augmentation injects only PE, and the resulting PE distribution remains close to that of the truth (Fig.~\ref{fig:KP_energy_GSonly_amplitude}).
Since the augmentation adds energy, this process is to be balanced by the additional dissipation of the KE, and this requires additional PE-into-KE energy conversion.
We confirmed that this conversion indeed happens and is achieved by more pronounced correlations (vector alignment) between the flow velocity and pressure gradient.
\ansD{PE-into-KE energy conversion seems to be incorrectly represented by the model due to the lack of
resolution; still the PE-into-KE route is more important and probably works better than the opposite one.}
Overall, conversion of PE into KE in the augmented solution is larger by an order of magnitude.
Viscous dissipation is about 3 times larger \ansC{than in the true solution} because of the excessive velocity values and gradients on mesoscales --- 
this suggests that augmentation of the tracer equations can be accompanied by increased eddy viscosity that should help to drain the excessive kinetic energy out of the system.
\ansC{However, more research is needed to understand what the best course of action is.}

%
%%%%%%%%%%%%%%%%%%% CONCLUSIONS AND DISCUSSION
%
\section{Conclusions and discussion}
In this work we have considered a comprehensive ocean model coupled with an atmospheric model in the North Atlantic configuration and characterized by different spatial resolutions.
We studied how augmentation of different equations of the coarse-grid model influences the large-scale circulation, with the particular focus on the Gulf Stream and its eastward extension.
We found that augmentation of only the momentum equation and the atmospheric model leads to insurmountable degradation of the tracer fields (temperature and salinity) and, therefore, requires additional augmentation of them.
\ansB{These findings also echo the results in~\citep{SGKBCA_2020} showing that unresolved temperature and salinity fields can produce significant errors in the density field, thus, affecting the momentum equation via the pressure gradient force.}

\ansD{When we correct the momentum equation at some time step, its predicted velocity becomes right.
However, velocity acting in the tracer equation remains uncorrected on this iteration.
Besides, the advection operator in the tracer equation generates its own errors, as it operates on the coarse grid.
The resulting error in this time step may seem to be small, but actually it is not.
One of the messages of our study is to show critical importance of even small errors in the tracer equations.}
Moreover, if only one of the tracer equations is augmented, then the solution becomes quickly contaminated with high-frequency waves and the large-scale flow structure becomes ruined.
Augmentation of the tracer equations only is sufficient for preserving the large-scale flow, whereas there is no need for parameterising the other oceanic equations and the atmospheric model. 
Moreover, the augmentation of the tracer equations can be accompanied by increased eddy viscosity to drain the excessive kinetic energy out of the system.
Note that this energy draining is opposite to the KE backscatter idea, whereas the crucially important PE injection can be interpreted as the ``PE backscatter''.
Overall, our findings suggest that \ansB{it may be sufficient for} parameterisations of comprehensive ocean models \ansB{to target only} the tracer equations.
\ansB{Note that if the goal is parameterisations based on closures then these suggestions are hardly valid as our approach cannot be used to close the model.
However, given the fact that data-driven parameteriations are picking up speed, our results suggest that parameterising tracers instead of momentum yields better results.}

\ansB{Our approach was deliberately taken to be the ``upper bound'', in the sense that we used complete information and targeted the full flow.
Using incomplete information about the eddy effects would be a useful future research program aiming to separate from each other more and less important aspects of the eddy effects.
Obviously, practical parameterizations can not handle everything and should focus on the most important aspects.
Second, although we targeted the full flow, we demonstrated that the most important statistical characteristics --- time mean circulation and stratification --- become unavoidably degraded without augmentation of the
dynamical tracers.
Considering second-moment statistics would be a natural extension of the analyses, but this goes beyond the scope of the present short communication.
}

Let us now have some critical discussion of the results and conclusions.
\ansC{Conceptually, it might be possible to achieve better results with first improving the temperature and salinity equations, but in practice development of successful parameterisation can be a lot more complicated due to many important nonlinear feedbacks in the system.
Moreover, our augmentation affects all scales whereas the classical parameterisation approach focuses on specific scales and physical processes~\citep{WHGSJ_2016} --- this difference needs to be recognized and taken into account.
Improvements in the North Atlantic model may come from improved representation of mesoscale dynamics or better resolved boundaries and bathymetry, or may be due to the improved vertical resolution, surface forcing, and better numerics.
Besides, the presented results may also be resolution-dependent (like Gent-McWilliams parameterization and kinetic energy backscatter) and domain-dependent, and exploring this is left for the future.   
}

We hope that our study will raise many new research questions, and some of them are at the surface.
\ansA{For example, we did not separate eddies from non-eddies and treated all motions nominally resolved on the $1/3^{\circ}$ grid as the large scales.
In the future it will be useful to introduce some spatial filtering for separating eddies from non-eddies (even though this separation is non-unique and raises many other issues) and to augment only the large-scale circulation, while monitoring what happens with the eddies.
Alternatively, one can augment only the eddies and see whether in this case their (under)resolved dynamics will be capable of augmenting the large scales.
The other useful future avenue is to understand by which process the added potential energy is converted into the kinetic energy, when only the tracer equations are augmented.
% Jamet et al. (2021) recently argued that the eddy-mean flow interaction in the separated Gulf Stream was dominated by barotropic processes.
}

\section{Acknowledgments}
The authors thank the Leverhulme Trust for the support of this work through grant RPG-2019-024 and Dr Quentin Jamet for helping us to set up the North Atlantic configuration. 
%Pavel Berloff was supported by the NERC grants NE/R011567/1 and NE/T002220/1, and by the Moscow Center for Fundamental and Applied Mathematics (supported by the Agreement 075-15-2019-1624 with the Ministry of Education and Science of the Russian Federation).



%% References


% \bibliographystyle{elsarticle-num}
\bibliographystyle{apalike}
\bibliography{refs}



\end{document}